\documentclass[twocolumn,showpacs,amsmath,amssymb,superscriptaddress,longbibliography]{revtex4-1}
\usepackage{amsmath,amssymb,color,latexsym,natbib,graphicx,subfigure}
\usepackage{dcolumn}
\usepackage{bm}
\usepackage{color}
\usepackage{ulem}
\def\be{\begin{equation}}
\def\ee{\end{equation}}
\def\ba{\begin{eqnarray}}
\def\ea{\end{eqnarray}}

\def\kpa{k_z}
\def\kpn{k_{\perp}}

\def\bbl{{\bf B_0}}
\providecommand\Ome{\boldsymbol{\Omega}}
\def\Omes{\Omega_0}
\def\uu{{\bf u}}
\def\bb{{\bf b}}
\def\ww{{\bf w}}

\def\kk{{\bf k}}
\def\zz{{\bf z}}

\def\ep{{\bf \hat{e}}_z}
\def\ek{{\bf \hat{e}}_k}

\def\ete{{\bf \hat{e}}_{\theta}}
\def\efi{{\bf \hat{e}}_{\Phi}}
\def\kkp{{\bf k}_{\perp}}

\def\hh{{\bf h^{\Lambda}_k}}
\def\Lp{\Lambda^{\prime}}
\def\Correction#1{{\textcolor{black}{#1}}}

\begin{document}
\preprint{1}

\title{Inverse cascade of hybrid helicity in $B\Omega$-MHD turbulence}

\author{M\'elissa D. Menu}
\affiliation{Laboratoire de Physique des Plasmas, Univ. Paris-Sud, Univiversit\'e Paris-Saclay, \'Ecole polytechnique, 
CNRS, Sorbonne Univ., Observatoire de Paris, F-91128 Palaiseau Cedex, France}
\affiliation{LERMA, Observatoire de Paris, PSL Research University, CNRS, Sorbonne Universit\'e, F-75005 Paris, France}
\email{melissa.menu@lpp.polytechnique.fr}

\author{S\'ebastien Galtier}
\affiliation{Laboratoire de Physique des Plasmas, Univ. Paris-Sud, Universit\'e Paris-Saclay,  Institut universitaire de France, \'Ecole polytechnique, 
CNRS, Sorbonne Univ., Observatoire de Paris, F-91128 Palaiseau Cedex, France}
\email{sebastien.galtier@lpp.polytechnique.fr}

\author{Ludovic Petitdemange}
\affiliation{LERMA, CNRS, Observatoire de Paris, PSL Research University, Sorbonne Universit\'e, F-75005 Paris, France}
\email{ludovic.petitdemange@lra.ens.fr}

\date{\today}

\begin{abstract}
We investigate the impact of a solid-body rotation $\Ome_0$ on the large-scale dynamics of an incompressible magnetohydrodynamic turbulent flow in 
presence of a background magnetic field $\bbl$ and at low Rossby number. Three-dimensional direct numerical simulations are performed in a periodic box, 
at unit magnetic Prandtl number and with a forcing at intermediate wavenumber $k_f=20$. When $\Ome_0$ is aligned with $\bbl$ (i.e. 
$\theta \equiv \widehat{\left(\Ome_{0}, \bbl \right)} = 0$), 
inverse transfer is found for the magnetic spectrum at $k<k_f$. This transfer is stronger when the forcing excites preferentially right-handed (rather than 
left-handed) fluctuations; it is smaller when $\theta>0$ and becomes weak when $\theta \ge 35^o$. 
These properties are understood as the consequence of an inverse cascade of hybrid helicity which is an inviscid/ideal invariant of this system when 
$\theta=0$. Hybrid helicity emerges, therefore, as a key element for understanding rotating dynamos. 
Implication of these findings on the origin of the alignment of the magnetic dipole with the rotation axis in planets and stars is discussed. 
\end{abstract}

\maketitle

%%%%%%%%%%%%%%%%%%%%%
\section{Introduction}

\begin{table*}
\caption{Main parameters of the simulations cited in the text. From left to right we find the name of the simulation, the amplitudes of the rotation rate and of the 
uniform magnetic field, the hypoviscosity, the normalized cross-correlation, the angle $\theta \equiv \widehat{\left(\Ome_{0}, \bbl \right)}$, the Rossby number, the
Reynolds number and the time at which the simulation has been stopped. All the simulations have been performed with a resolution $N=256^3$ for a box size 
$L=2\pi$, a reduced magnetic helicity $\sigma_m = 0.5$ and a viscosity $\nu^+ = \eta^+ = 2.10^{-3}$.}
\label{table1}
\begin{ruledtabular}
\begin{tabular}{lcccccccc}
Simulation & $\Omega_0$ & $B_0$ & $\nu^-=\eta^-$ & $\sigma_c$ & $\theta$ & $Ro$ ($10^{-3}$) & $Re$ & $t_f$\\
{\bf 20L} & $20$ & $1$ & $0$ & $0.5$ &$0$ & $1.54$ & $1215$ & $69$\\
{\bf 20R} & $20$ & $1$ & $0$ & $-0.5$ &$0$ & $1.5$ & $1182$ & $69$\\
{\bf 0R} & $0$ & $1$ & $0$ & $-0.5$ &$0$ & $\infty$ & $1344$ & $62$\\
{\bf 20R\_B0} & $20$ & $0$ & $0$ & $-0.5$ &$0$ & $1.45$ & $1147$ & $97$\\
{$\bf 20R_0$} & $20$ & $1$ & $2.10^{-2}$ & $-0.5$ & $0$ & $1.49$ & $1176$ & $133$\\
{$\bf 20R_{25}$} & $20$ & $1$ & $2.10^{-2}$ & $-0.5$ & $25$ & $1.49$ & $1175$ & $105$\\
{$\bf 20R_{35}$} & $20$ & $1$ & $2.10^{-2}$ & $-0.5$ & $35$ & $1.47$ & $1163$ & $113$\\
{$\bf 20R_{45}$} & $20$ & $1$ & $2.10^{-2}$ & $-0.5$ & $45$ & $1.49$ & $1178$ & $117$\\
{$\bf 20R_{90}$} & $20$ & $1$ & $2.10^{-2}$ & $-0.5$ & $90$ & $1.57$ & $1241$ & $108$\\
\end{tabular}
\end{ruledtabular}
\end{table*}

The emergence of large-scale magnetic fields in various astrophysical objects (like planets, stars, accretion discs or galaxies) is mainly attributed 
to a dynamo mechanism based on the turbulent motions of a conducting fluid described by magnetohydrodynamics (MHD) 
\cite{Moffatt72,Brandenburg05,Kulsrud08,Gcup16,Moutou17}. Because the magnetic flux is conserved in ideal MHD, the stretching of magnetic field 
lines by the conducting fluid can amplify magnetic fluctuations at small scales. It is thought that these turbulent fluctuations are then transported to 
large-scales {\it via} an inverse cascade of magnetic helicity \cite{Frisch75,pouquet76,Pouquet78,Meneguzzi81,Alexakis06}, which is an ideal invariant 
of three-dimensional (3D) MHD \cite{Woltjer1958}. The presence of inverse transfer of magnetic energy in absence of magnetic helicity is 
also possible as pointed out in \cite{Brandenburg15} (see also \cite{Gilbert88,Urpin02,Tobias13}). This transfer is, however, weaker than the one 
found with magnetic helicity and could be explained e.g. by the form of the initial spectrum in the sub-inertial range \cite{Olesen97}. 

Strictly speaking, direct and inverse cascades are expected only for quantities which are invariant of a system in the non-dissipative case, 
whatever the turbulence regime (strong or weak) \cite{frisch95,Nazarenko11,Banerjee14,Seshasayanan14,Alex18}. In 3D incompressible MHD, 
such invariants are the total energy $E$, the cross-correlation between the velocity and the magnetic field $H_c$, and the magnetic helicity $H_m$ 
\cite{Pouquet93}. There are many studies devoted to the scaling of the total energy spectrum for which the answer is not unique 
\cite{Kraichnan65,Goldreich95,galtier00,GMP05,Boldyrev2006,Mininni07,Lee10,beresnyak14}. Much less is known about the magnetic helicity 
while its importance is recognized e.g. in solar physics where $H_m$ can be measured in coronal mass ejections \cite{priest} or in the solar wind 
\cite{matthaeus82}. 
Recently, several direct numerical simulations have been devoted to the study of the magnetic helicity cascade \cite{Alexakis06,Muller12,Linkmann16} 
(see also \cite{brandenburg01} for compressible MHD). In particular, it is shown that the inverse cascade becomes nonlocal in wavenumber space 
when condensation takes place at the largest scale of the system. Under some conditions, a direct cascade of $H_m$ can also be found as a finite 
magnetic Reynolds number effect \cite{Linkmann17}. 

The introduction of a uniform magnetic field $\bbl$ or the Coriolis force with a uniform rotating rate $\Ome_0$ reduces the number of inviscid/ideal 
invariants in 3D incompressible MHD. In the first case, $H_m$ is no longer conserved while in the second it is $H_c$. When both effects are present, 
(situation called hereafter, $B\Omega$-MHD) the total energy remains the only invariant of the system, except if $\bbl$ and $\Ome_0$ 
are aligned: 
in this case, there is a second invariant called hybrid helicity $H_h$, which is a combination of $H_c$ and $H_m$ \cite{Shebalin06}. While analytical 
results have been obtained recently for weak $B\Omega$-MHD turbulence \cite{Galtier14} with some predictions about the hybrid helicity spectrum, 
no detailed numerical study has been done in the strong or weak wave turbulence regime 
\Correction{(see, however, the recent study by \cite{Bell19})}.
$B\Omega$-MHD turbulence is, however, a relevant model for studying rotating dynamos like in stars and planets which are often characterized by 
a magnetic dipole closely aligned with the rotation axis. The reason of this alignment is still unclear and need further investigations. Because of the 
complexity of the problem, only few physical ingredients are generally included in the modeling (see eg. \cite{Reshetnyak08,Petitdemange18}). For 
example, we may investigate this problem by including a large-scale magnetic field $\bbl$ and/or a solid-body (instead of differential) rotation 
$\Ome_0$ (see eg. \cite{Favier12,Seshasayanan17}). 

In this article, we present a set of 3D direct numerical simulations of $B\Omega$-MHD turbulence at unit magnetic Prandtl number and low Rossby number. 
The investigation is focused on the large-scale dynamics (scales larger than the forcing scale). 
In section \ref{sec2} we present the governing equations and the numerical setup. Section \ref{sec3} is devoted to the numerical results. 
When the angle $\theta \equiv \widehat{\left(\Ome_{0}, \bbl \right)}$ between $\Ome_0$ and $\bbl$ is null, we show 
that the magnetic spectrum exhibits a significant inverse transfer which is reduced when $\theta>0$ to become negligible for $\theta \ge 35^o$. We also 
show that this transfer is stronger when the forcing excites preferentially right-handed (rather than left-handed) fluctuations. We explain why these 
properties can be interpreted as the consequence of an inverse cascade of $H_h$, which appears as a key element to understand rotating dynamos. 
Finally, in section \ref{sec4} we present a conclusion. 

%%%%%%%%%%%%%%%%%%%%%
\section{Governing equations}\label{sec2}
%%%%%%%%%%%%%%%%%%%%%

The equations governing incompressible $B\Omega$-MHD can be written as
\ba
\frac{\partial \uu}{\partial t} &=& - \nabla P + \uu \times (\ww + 2 \Ome_0) + (\nabla \times \bb) \times (\bb + \bbl) \nonumber \\
&&+ \nu^+ \nabla^2 \uu + \nu^- \nabla^{-2} \uu \, , \label{rmhd1} \\
\frac{\partial \bb}{\partial t} &=& \nabla \times (\uu \times (\bb+\bbl))  
+ \eta^+ \nabla^2 \bb + \eta^- \nabla^{-2} \bb \, , \label{rmhd2} \\
&& \nabla \cdot \uu = 0 \, , \label{rmhd3} \\
&& \nabla \cdot \bb = 0 \, , \label{rmhd4}
\ea
with $\uu$ the velocity, $P$ a generalized pressure, $\ww=\nabla \times \uu$ the vorticity and $\bb$ the normalized magnetic field. 
$\nu^+$, $\eta^+$ and $\nu^-$, $\eta^-$ are small-scale and large-scale dissipation coefficients, respectively. 
We can easily check that $H_c$ and $H_m$ are not conserved in $B\Omega$-MHD since we obtain from Eqs. (\ref{rmhd1})--(\ref{rmhd2}), 
with $\nu^+=\nu^-=\eta^+=\eta^-=0$,
\ba
\Correction{d H_c \over d t} &\equiv& {d \over d t} \int \frac{\uu \cdot \bb}{2} \, d\,{\cal V} 
= \Ome_0 \cdot \int (\bb \times \uu) \, d\,{\cal V} \, , \\
\Correction{d  H_m \over d t} &\equiv& {d \over d t} \int \frac{{\bf a} \cdot \bb}{2} \, d\,{\cal V} 
= \bbl \cdot \int (\bb \times \uu) \, d\,{\cal V} \, , 
\ea
where ${\bf a}$ is the vector potential ($\bb = \nabla \times {\bf a}$) and ${\cal V}$ a volume of integration. The hybrid helicity $H_h \equiv H_m/d - H_{c}$ 
with $d \equiv B_0 / \Omes$ is, however, conserved when $\Ome_0$ and $\bbl$ are aligned (this property is checked numerically but not shown). 
$d$ is called the magneto-inertial length and gives a scale of reference to measure the relative importance of the Coriolis force on the Lorentz force. 

The linear solution of MHD is modified by the presence of the Coriolis force; the dispersion relation is \cite{Finlay,Galtier14}
\be
\omega = \frac{s k_z \Omes}{k} \left( -s\Lambda + \sqrt{1+k^2d^2} \right) \, , \label{dispz}
\ee
with $k$ the wavenumber, $k_z$ the wavenumber component along $\bbl$ (here, we assume that $\Ome_0$ and $\bbl$ are parallel), 
$s=\pm 1$ the directional polarity ($sk_z \ge 0$) and $\Lambda=\pm 1$ the wave polarization. The magnetostrophic branch ($\Lambda s=1$) and the inertial branch 
($\Lambda s=-1$) correspond to the right (R) and left (L) circular polarizations, respectively. There are well separated when $kd\ll1$ and tends to the 
Alfv\'en branch when $kd \gg 1$, also the condition $kd=1$ appears as a critical value. 
As shown in \cite{Galtier14}, \Correction{the polarization $P$ may be defined as} 
$\Correction{P \equiv} \sigma_m \sigma_c = - s \Lambda$, with $\sigma_m$ the reduced magnetic helicity and $\sigma_c$ the reduced cross-correlation
\ba
\sigma_m &=& \frac{\hat{\bf a} \cdot \hat{\bb}^{*} + \hat{\bf a}^* \cdot \hat{\bb} }{2 \vert \hat{\bf a} \vert \vert \hat{\bb} \vert} \, , \\ 
\sigma_c &=& \frac{\hat{\uu} \cdot \hat{\bb}^*+\hat{\uu}^* \cdot \hat{\bb}}{2 \vert \hat\uu \vert \vert \hat\bb \vert} \, ,
\ea
where $\hat{}$ means the Fourier transform and $*$ the complex conjugate. 
\Correction{By extension, in our numerical study we define the R and L fluctuations for which we have, respectively, $P<0$ and $P>0$.}
Finally note that the polarization property is lost when $\Omes=0$, while when $B_0=0$ we end up with a L-polarized wave, giving an asymmetric character to the dispersion relation (\ref{dispz}). 

Equations (\ref{rmhd1})--(\ref{rmhd3}) are computed using a pseudo-spectral solver called TURBO \cite{Meyrand12,Meyrand16}. The simulation domain 
is a triply periodic cube discretized by $N^3$ collocation points. A unit magnetic Prandtl number is taken with $\nu^+=\eta^+$; we also take $\nu^-=\eta^-$. 
The vector $\bbl$ is fixed along the z-direction while $\Ome_0$ may be tilted with an angle $\theta$ in such a way that for $\theta=0$ it is also along the 
z-direction. The nonlinear terms are partially de-aliased using a phase-shift method. This system is forced in the Fourier space: kinetic and magnetic 
energy spectra are excited at wavenumbers $19 \le k_f \le 21$ with a rate of injection $\epsilon_u$ and $\epsilon_b$, respectively, while there is no injection 
of kinetic helicity \cite{Teaca11}. We take $\epsilon_u=\epsilon_b=0.2$ for all simulations. Magnetic helicity and cross-correlation are also 
injected at $k_f$ with a reduced rate $\sigma_m$ and $\sigma_c$, respectively. We take $\sigma_m=0.5$, then the sign of $\sigma_c$ will determine the 
polarization (left or right). 
\Correction{In real systems this type of polarized forcing may find its origin in the excitation of magnetostrophic or inertial waves preferentially (see eg. \cite{LeReun17}).} 
Note that the simulations have been stopped at a time $t_f \gg t_{NL} \sim 1$, where $t_{NL}$ is the nonlinear time, i.e. the time 
needed to form the small-scale ($k>k_f$) spectra. Therefore, the dynamics that we investigate is relatively slow and requires a long numerical computation. 

A summary of the different simulations is given in Table \ref{table1}. In particular, the choice of $\Omega_0$ and $B_{0}$ is made to keep $k_f d=1$. 
Simulations have been computed to obtain a sufficiently large steady state window 
%at least two decades of steady state 
for the kinetic energy (with a steady state that starts at $t_{steady} \sim 40$ for cases 20L, 20R, 0R and $t_{steady}\sim80$ for all the others). 
The Rossby, $Ro = U/(2\Omega_0 L)$, and Reynolds, $Re = UL/\nu^+$, numbers are calculated from the root mean square value of the velocity field averaged 
over the entire volume $V$ of the numerical box and time-averaged for $t_{steady} \le t \le t_f$: $U = \langle \langle \uu \cdot \uu \rangle_V \rangle_t ^{1/2}$.

%%%%%%%%%%%%%%%%%%%%%
\section{Results}\label{sec3}
%%%%%%%%%%%%%%%%%%%%%

\subsection{Impact of the circular polarizations}\label{Polar}
\begin{figure}
\includegraphics[width=1\linewidth]{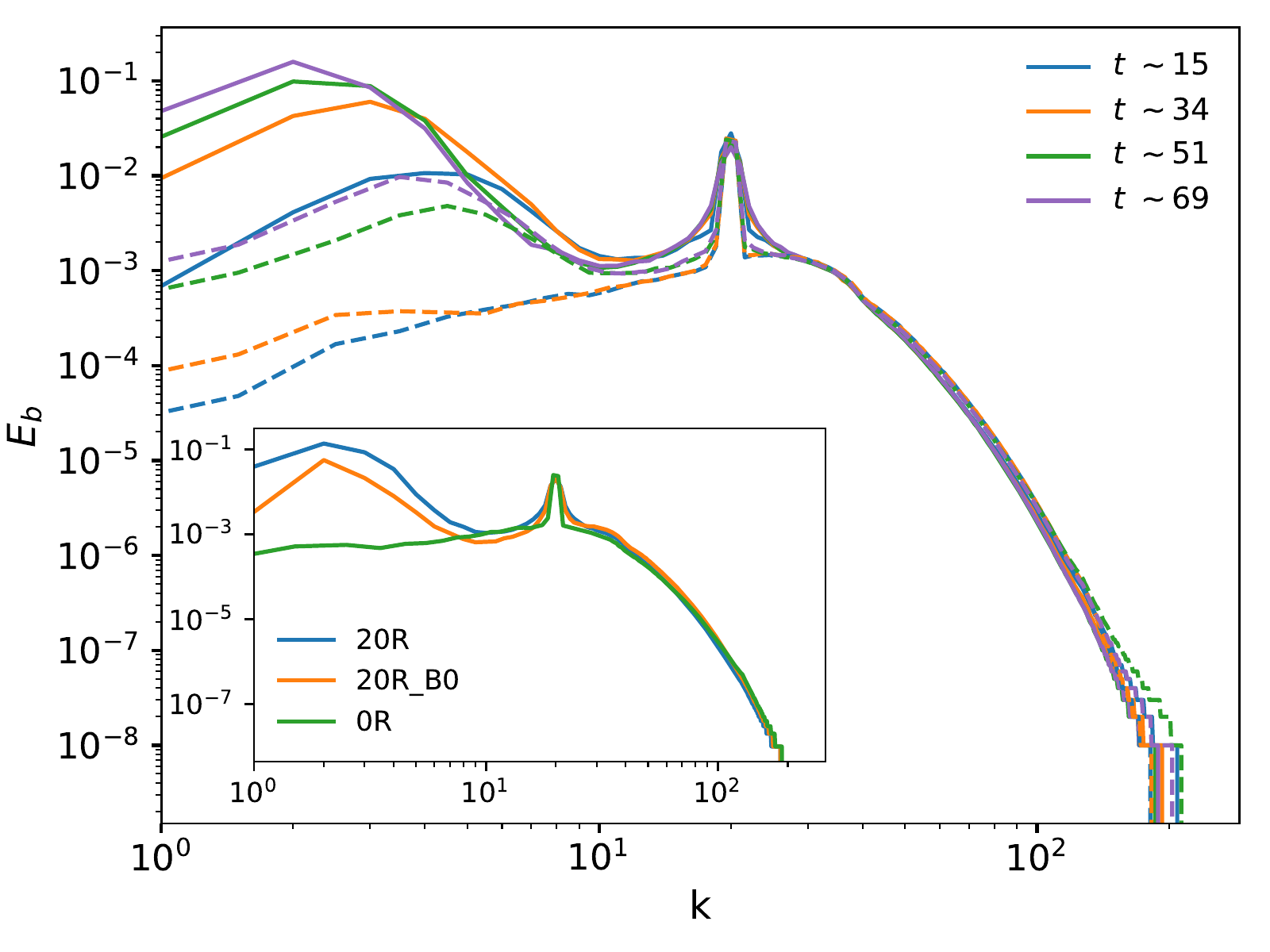}
\caption{Temporal evolution of the magnetic spectrum for simulations 20R (solid line) and 20L (dashed line). The parameters of these simulations can be found in Table \ref{table1}. Colors indicate the approximate times at which the spectra have been computed. Note that simulation 20R is stopped before the formation of a \Correction{condensate}. Insert: spectrum of case 20R at $t\sim62$ compared with two reference cases: without rotation (case 0R) and without magnetic 
field (case 20R\_B0).}
\label{Fig1}
\end{figure}
$B\Omega$-MHD turbulence is characterized by two types of fluctuations (R and L). We start our analysis by studying the impact on the large-scale 
dynamics of a forcing which excites preferentially the R (simulation 20R) or the L (simulation 20L) fluctuations. 
Both magnetic $E_b$ and kinetic $E_u$ energies spectra have been calculated to diagnose the dynamics, however, $E_u$ is nearly constant for these 
two cases. The behavior of the kinetic energy will be briefly discussed in section \ref{Angles}. Fig. \ref{Fig1} shows the results with 
the time evolution of the magnetic spectrum. The plots are given for approximately the same times. The simulation 20R is stopped 
before the formation of a \Correction{condensate} at low wave numbers which may have an impact on the dynamics (finite box effect). In both cases inverse transfers of 
magnetic energy are found for $k<k_f$, however, we clearly see that the efficiency of the transfer is greater when the R-fluctuations are preferentially excited: the magnetic energy transfer to large scales occurring in case 20L (dashed line) is considerably less efficient than in case 20R as the maximum value reached by 
the magnetic energy at the final time $t\sim 69$ differs by more than an order of magnitude. This difference can be understood by using wave turbulence 
arguments: the dynamics of the R-fluctuations is mainly driven by the magnetic field while it is mainly driven by the velocity field for the L-fluctuations 
\cite{Galtier14}. Therefore, simulation 20R (solid line) strengthens a dynamics driven by the magnetic field.
The efficiency of this transfer is also compared (see insert) to a rotating case without magnetic field (simulation 20R\_B0). The same behavior is observed, 
however, we see that adding a mean magnetic field to the strong rotation enhances slightly the inverse transfer of magnetic energy. The situation is quite 
different when we remove the rotation (simulation 0R): in this case the fluctuations are not circularly polarized, the large-scale magnetic spectrum is flat and 
does not evolve very much. 

\begin{figure}
\includegraphics[width=1\linewidth]{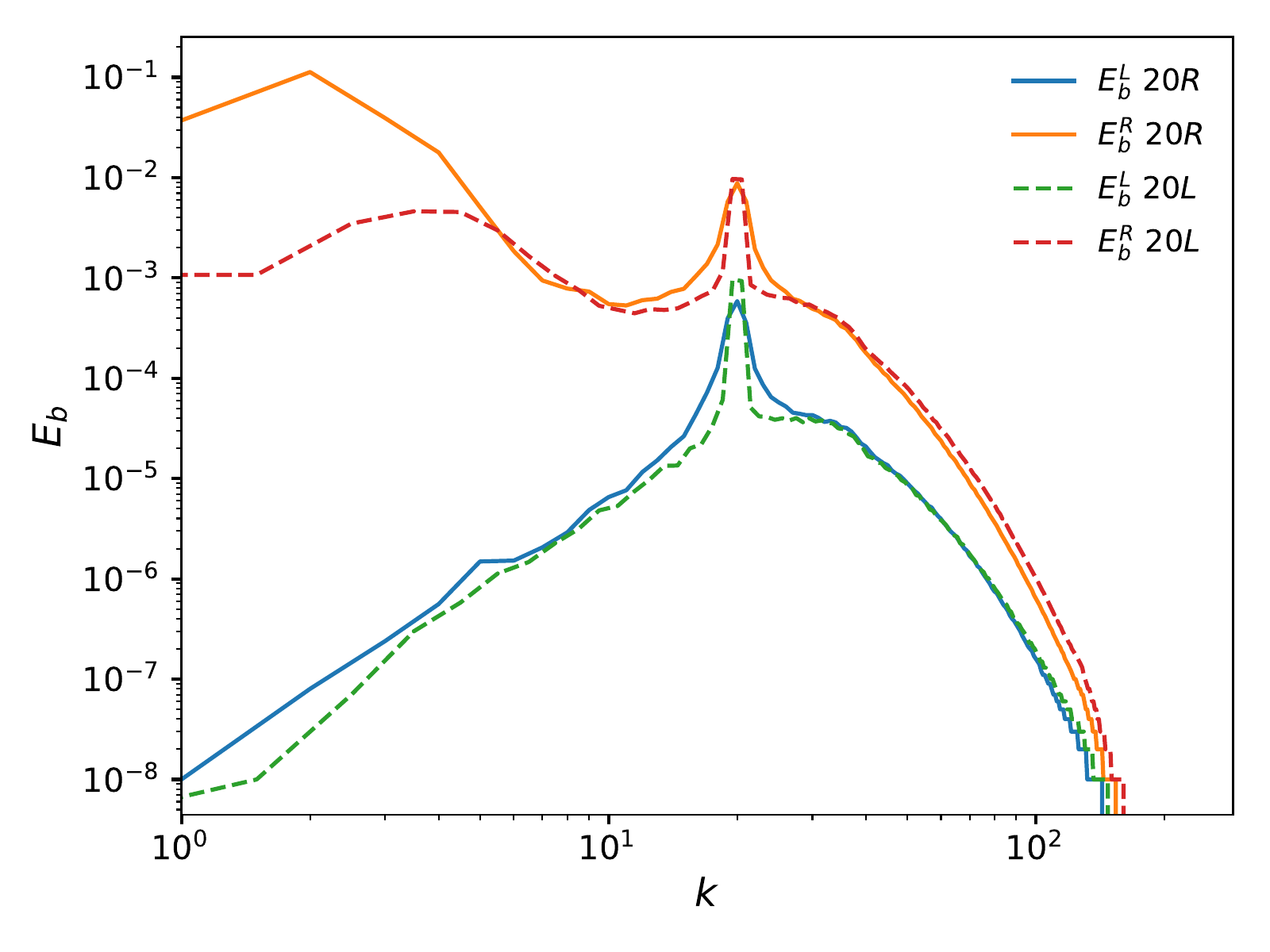}
\caption{Magnetic spectra at $t\sim69$ for simulations 20R (solid line) and 20L (dashed line) decomposed into L ($E^L_b$) and R ($E^R_b$) fluctuations. 
This decomposition highlights the relative importance of each type of fluctuations in a single simulation.}
\label{Fig2}
\end{figure}
Fig. \ref{Fig2} confirms the first picture by showing the spectra of Fig. \ref{Fig1} but only for the final times and decomposed into L- and R-fluctuations 
(the decomposition is discussed in \cite{Galtier14}; see also Appendix \ref{sec5}). For both simulations we see that the inverse cascade involves mainly the 
R-fluctuations and these fluctuations are larger for simulation 20R: the R-fluctuations drive the mechanism of inverse transfer in both simulations, whereas the L-fluctuations are significantly smaller at large scales. 
\Correction{This difference can be explained as the condition $k_f d=1$ leads to a magnetostrophic regime (R polarization) at large scales ($k<k_{f}$). As expected, the efficiency of a L-type forcing (dashed line on Fig. \ref{Fig2}) to drive an inverse cascade of magnetic energy is significantly weaker than for a R-type forcing (solid line on Fig. \ref{Fig2}).}
Note that a similar analysis was performed for studying Hall MHD turbulence where a different behavior was also found for the L and R magnetic 
fluctuations spectra \cite{Meyrand12}. 

\begin{figure}
\includegraphics[width=1\linewidth]{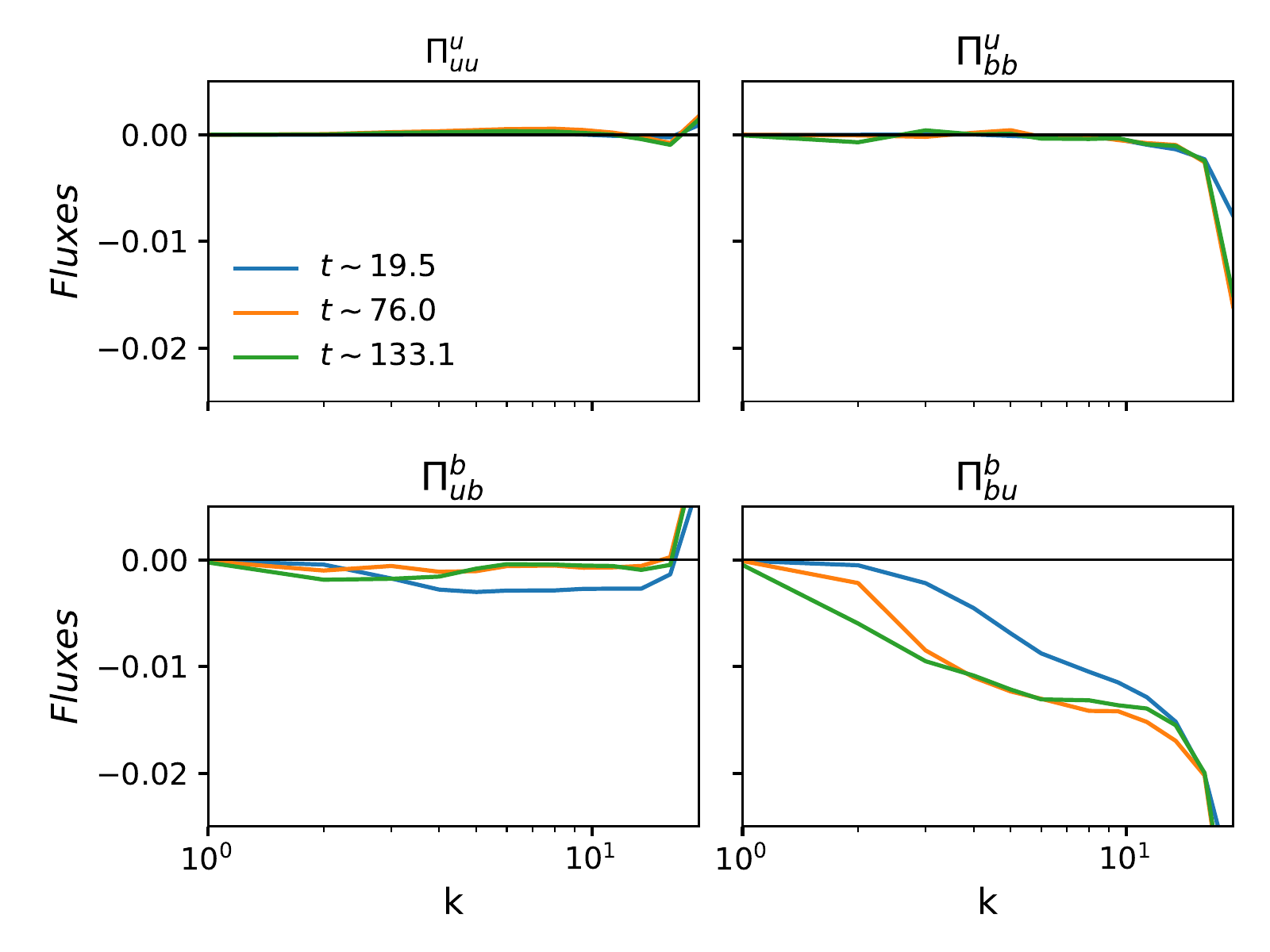}
\caption{Spectral flux (in logarithmic-linear coordinates) associated with the four different nonlinear terms of $B\Omega$-MHD (simulation 20R) for three
typical times. The main contribution exhibiting a significant negative flux at large scales ($k<k_f$) comes from $\Pi^b_{bu}$.}
\label{Fig3}
\end{figure}
In Fig. \ref{Fig3} we plot in the spectral space (for simulation 20R) the contributions of the four different terms of $B\Omega$-MHD to the total energy flux 
\Correction{as derived by \citet{Dar01} and \citet{Verma04}} for $k<k_f$ only. We used the notation $\Pi^X_{YZ}(k)$ for a flux \Correction{from inside the shell $k$ of the field X to outside the shell $k$ of field Z} via field Y.
By definition \cite{frisch95,Dar01,Verma04,Sesha16,Verma17,Meyrand18}, we have:
\ba
\Pi^{u}_{uu} (k) &=& \uu_{k}^{<} \cdot(\uu \cdot \nabla \Correction{\uu_{k}^{>} }) \, , \\
\Pi^{u}_{bb} (k) &=& - \uu_{k}^{<} \cdot(\bb \cdot \nabla \Correction{\bb_{k}^{>} } ) \, , \\
\Pi^{b}_{ub} (k) &=& \bb_{k}^{<} \cdot(\uu \cdot \nabla \Correction{\bb_{k}^{>} }) \, , \\
\Pi^{b}_{bu} (k) &=& - \bb_{k}^{<} \cdot(\bb \cdot \nabla \Correction{\uu_{k}^{>} } ) \, ,
\ea
where $\uu_{k}^{<}$ is the filtered velocity (or magnetic field $\bb_{k}^{<}$) so that only the modes $\vert \kk \vert < k$ are being kept. While the contribution 
from the advection term $\Pi^{u}_{uu}$ has the smallest amplitude, we see that the main contribution to the negative flux at large scales ($k<k_f$) comes from 
$\Pi^{b}_{bu}$. Although the range of scales is narrow, a plateau seems to emerge with time. We also see that there is a non negligible contribution of flux 
$\Pi^{b}_{ub}$ with a negative value. These two fluxes come from the induction equation, which is consistent with our interpretation (a dynamics dominated by 
the magnetic field).  
The fluxes at $k_f > 20$ (not presented) have a classical positive and decreasing shape from the forcing wave numbers to the 
dissipative scales, signature of a direct cascade.

\subsection{Impact of a tilted rotation axis}\label{Angles}
\begin{figure}
\includegraphics[width=1\linewidth]{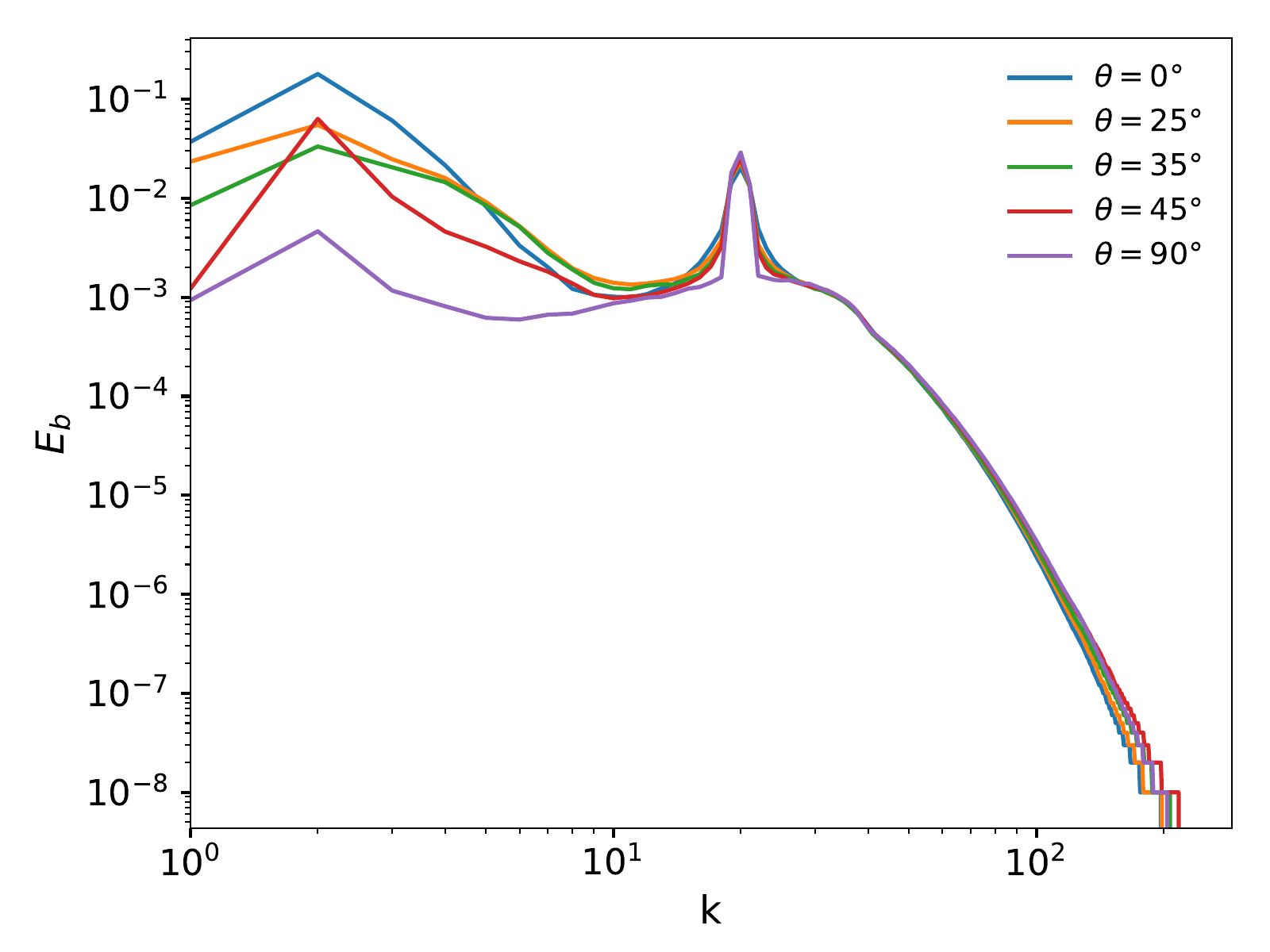}
\caption{Magnetic spectra for situations where the angles $\theta=0^o, 25^o, 35^o, 45^o$ and $ 90^o$ (simulations 20R$_0$ to 20R$_{90}$). 
The spectra are plotted approximately at the same time $t\sim 105$. Hypoviscosity ($\nu^-$) has been added to avoid a condensation at small $k$. 
At large scales, a similar shape is observed for $\theta \leq 35^o$ whereas the behavior seems different when $\theta \geq 45^o$.}
\label{Fig4}
\end{figure}

The hybrid helicity is an invariant of non-dissipative $B\Omega$-MHD only when $\theta=0^o$. Here, we study the impact of this angle on the large-scale dynamics. 
For this study, a hypoviscosity term has been added ($\nu_- \neq 0$; see Table \ref{table1}) to avoid the condensation observed in section \ref{Polar} and the finite 
box effects at small wavenumber. We will assume for the moment that $H_h$ is mainly driven by the magnetic helicity (see Fig. \ref{Fig7} for a
justification). Fig. \ref{Fig4} shows the results for five different angles. The same forcing as in simulation 20R is applied. A significant 
decrease of the inverse transfer is observed when the angle $\theta$ increases. For $\theta=90^o$ the transfer can be qualified as negligible. This 
property of $B\Omega$-MHD turbulence can be interpreted as the direct consequence of the non conservation of $H_h$: the large-scale dynamics 
observed for $\theta=0^o$ is explained by the inverse cascade of $H_h$ which decreases when $\theta > 0^o$. Whereas from a theoretical 
point of view we expect the absence of inverse cascade as soon as $\theta>0^o$, Fig. \ref{Fig4} reveals the existence of a gradual decrease of this
cascade. Moreover, the large scale behavior differs when $\theta \ge 45^o$ with the presence of a significant peak at wavenumber $k=2$ and a curve 
instead of power law for $k>2$. Further analysis reveals that this behavior seems correlated with that of $E_u$.  

\begin{figure}
\includegraphics[width=1\linewidth]{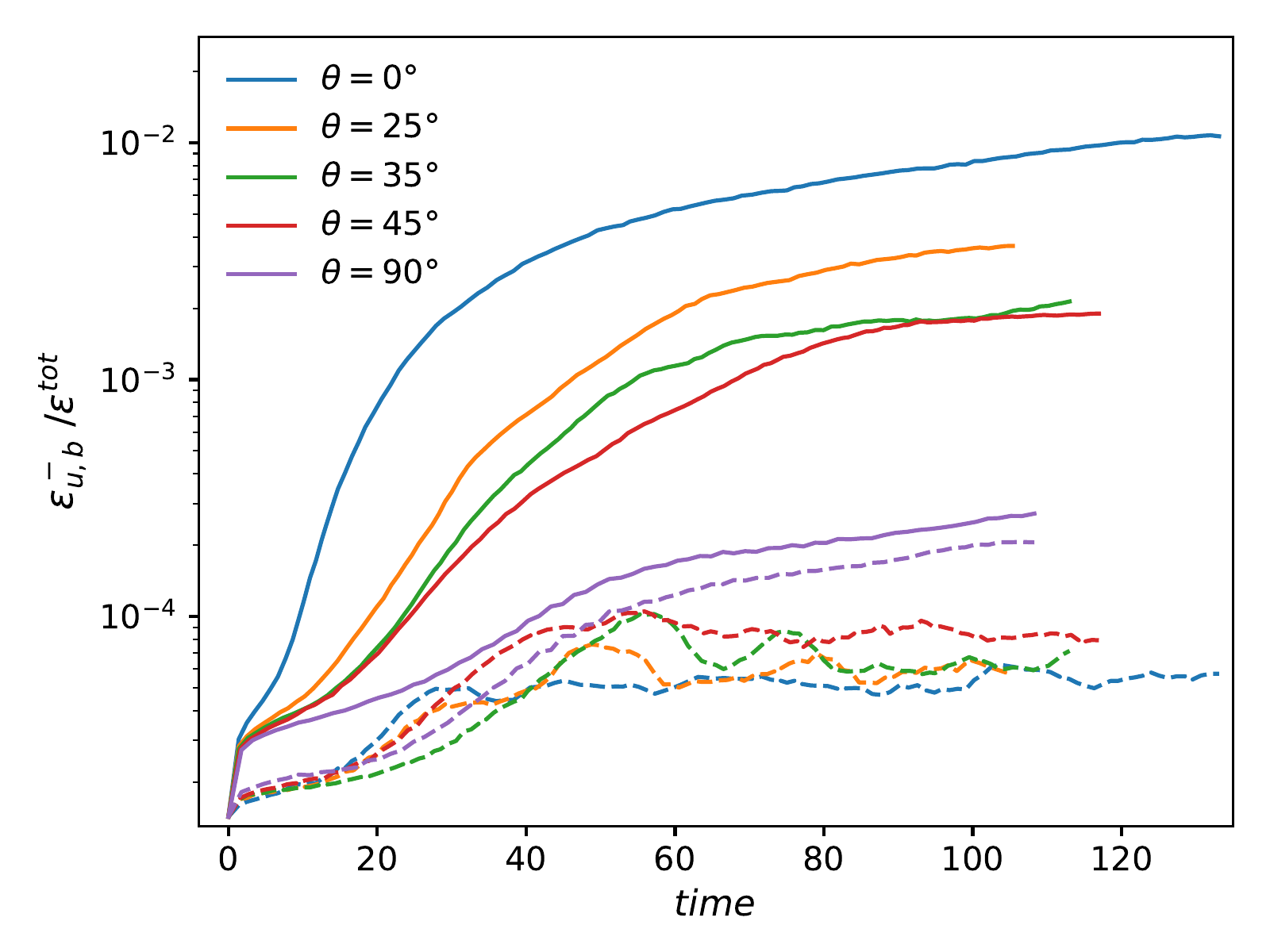}
\caption{Time evolution (in linear-logarithmic scales) of the normalized large-scale dissipation rate of kinetic energy $\epsilon^-_u$ (dashed lines) and 
magnetic energy $\epsilon^-_b$ (solid lines) for angles $\theta= 0^o$,  $25^o$, $35^o$, $45^o$ and  $90^o$. $\epsilon^{tot} \equiv \epsilon^+ + \epsilon^-$ 
is the total dissipation rate. For $\theta = 0^o$ the large scale dissipation rate is dominated by the magnetic contribution, while for $\theta = 90^o$ the contribution of each field is similar.}
\label{Fig5}
\end{figure}
A better way to quantify this evolution is to measure the dissipation rate of energy at small and large scales  
\be
\epsilon^\pm = \nu^\pm \sum_{\kk \neq {\bf 0}} k^{\pm2} ( \vert \hat{\uu} \vert^2 + \vert \hat{\bb} \vert^2 ) = \epsilon^\pm_u + \epsilon^\pm_b \, . 
\ee
In particular, $\epsilon^-_{u,b}/(\epsilon^+ + \epsilon^-)$ provides a measure of the strength of the inverse cascade \cite{Seshasayanan14}. 
Note that this measure does not require a mechanism of inverse cascade driven by the total energy. Fig. \ref{Fig5} displays the result for five angles. 
For $\theta=0^o$ we see that the fast growth of the large-scale magnetic dissipation observed initially is followed by a phase of slow growth meaning that 
the stationary state is only reached approximately. Interestingly the value obtained at the final time of the simulation is around $10^{-2}$, 
which means that most of the magnetic energy flux goes to small-scales, a property expected because of the direct energy cascade. 
The comparison with the other angles reveals a significant decrease of the large-scale magnetic dissipation and a slight increase of the large-scale kinetic 
dissipation. For $90^o$ an equipartition of the dissipation rates is almost reached. In this case the magnetic and kinetic energy spectra become very 
close (not shown). It is interesting to note that this tendency to the equipartition for $\theta = 90^o$ can be predicted already at the level of a linear 
analysis \cite{Sahli17}. In conclusion, this new diagnostic confirms the analysis made from Fig. \ref{Fig4} but in addition we can claim that the strength of 
the inverse cascade becomes significantly weaker (about an order of magnitude) when $\theta \ge 35^o$. 

\begin{figure}
\includegraphics[width=0.75\linewidth]{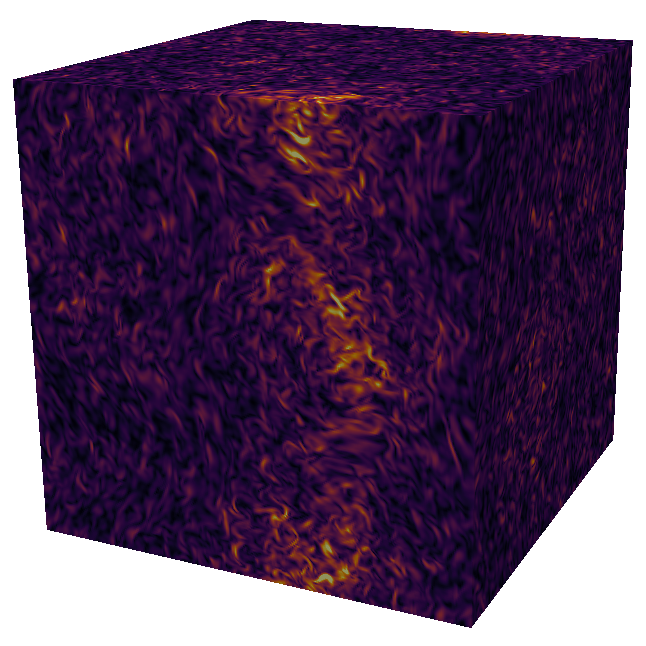}
\includegraphics[height=0.75\linewidth]{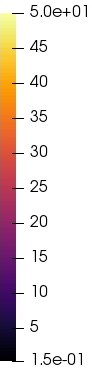}
\caption{\Correction{Amplitude of the current density for case $20R_0$ at $t \sim 105$. The magnetic field is along $\zz$, corresponding to the vertical axis on this representation. Despite strong rotation, no strong anisotropy is observed.}}
\label{Visu}
\end{figure}

\Correction{The amplitude of the current density is shown on Fig. \ref{Visu} for the reference case $\theta=0^o$ in 3D-space with the vertical axis corresponding to $\zz$. Despite the strong rotation imposed, there is no evidence for a strong anisotropy along the $\Omega_0$ direction. Especially no columnar structures leading to a quasi-2D turbulence are observed like for a purely rotational case (see eg. \cite{Sharma18}) or a purely magnetic case (see eg. \cite{Sundar17}) where, however, hypoviscosity/hyporesistivity was not introduced. By comparing their results to our similar cases (respectively 20R\_B0 and 0R), this quasi-2D behavior is not observed neither. 
This difference can be explained by different parameter ranges and also by a wave-type forcing which may prevent the formation of coherent structures. 
Vorticity field for the same simulation is pretty similar and therefore not shown. Note that no structures are observed when the rotation axis is tilted.}

\begin{figure}
\includegraphics[width=1\linewidth]{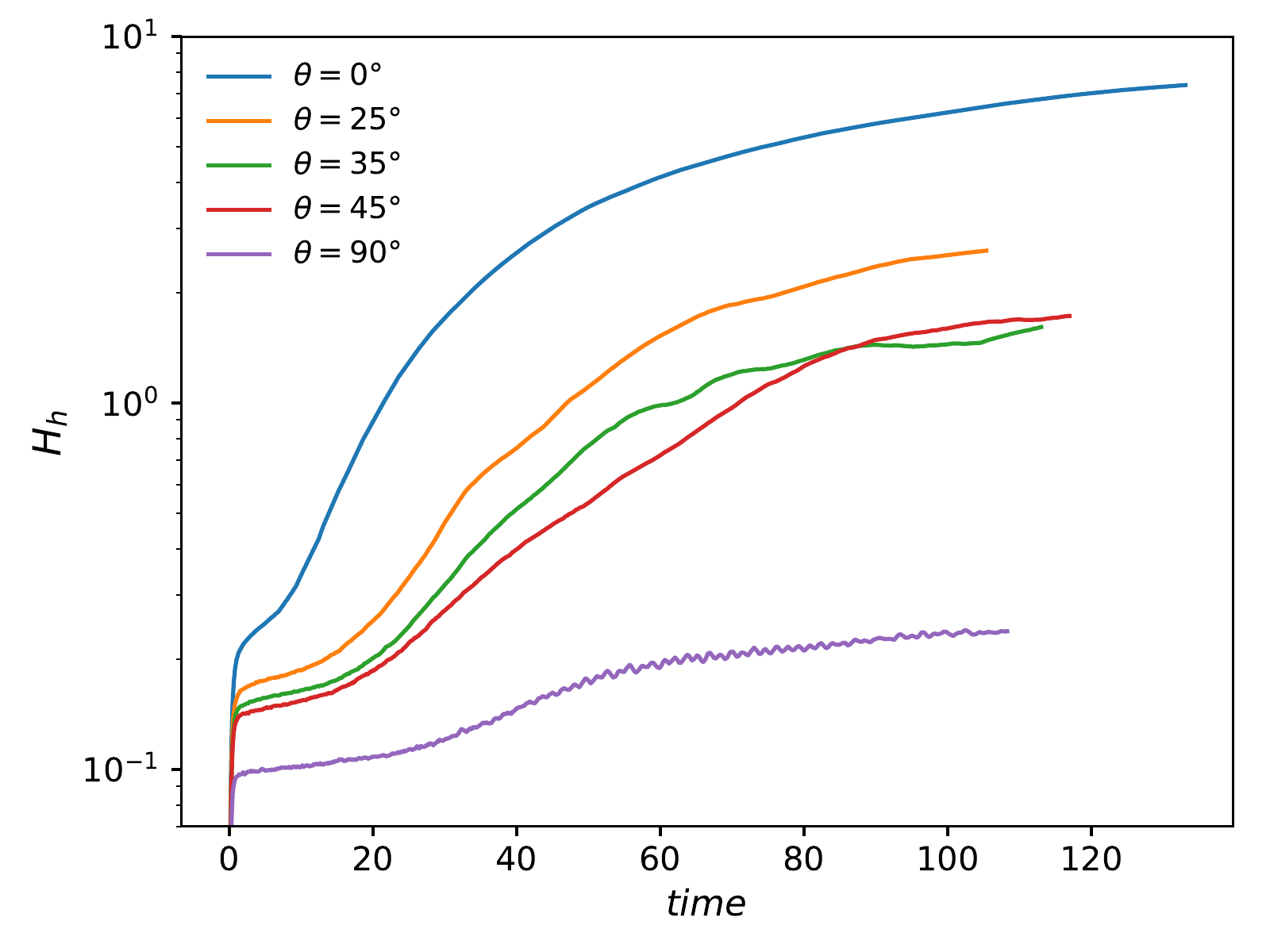}
\caption{Time evolution (in linear-logarithmic scales) of the hybrid helicity $H_h$ for the five simulations with different values of $\theta$. 
Note that the particular behaviors of case $20R_{35}$ and $20R_{45}$ which cross around $t=90$ seem to have different origins (see Fig. \ref{Fig4}). 
As expected, $H_h$ is much smaller for the orthogonal configuration $20R_{90}$.}
\label{Fig6}
\end{figure}
Fig. \ref{Fig6} shows the time evolution of the hybrid helicity with respect to $\theta$. We do not expect the conservation of this quantity in this case since 
an external forcing is applied. However, a stationary state may be reached in presence of hypoviscosity because of the balance between forcing and dissipation. 
Although the final times $t_f$ of the simulations are not exactly the 
same we see a general tendency with an accumulation of hybrid helicity into the system as a consequence of the inverse cascade. More than an order 
of magnitude of difference is found at $t_f$ between angles $\theta=0^o$ and $90^o$. 
The figure also shows that a stationary state is reached only approximately. 
Finally, note that the curves at $\theta=35^o$ and $45^o$ intersect around $t=90$. This observation has to be compared with the spectral behavior 
found in Fig.\,\ref{Fig4} at wavenumber $k=2$ to understand that the large scale repartition of energy is different. 

\begin{figure}
\includegraphics[width=1.\linewidth]{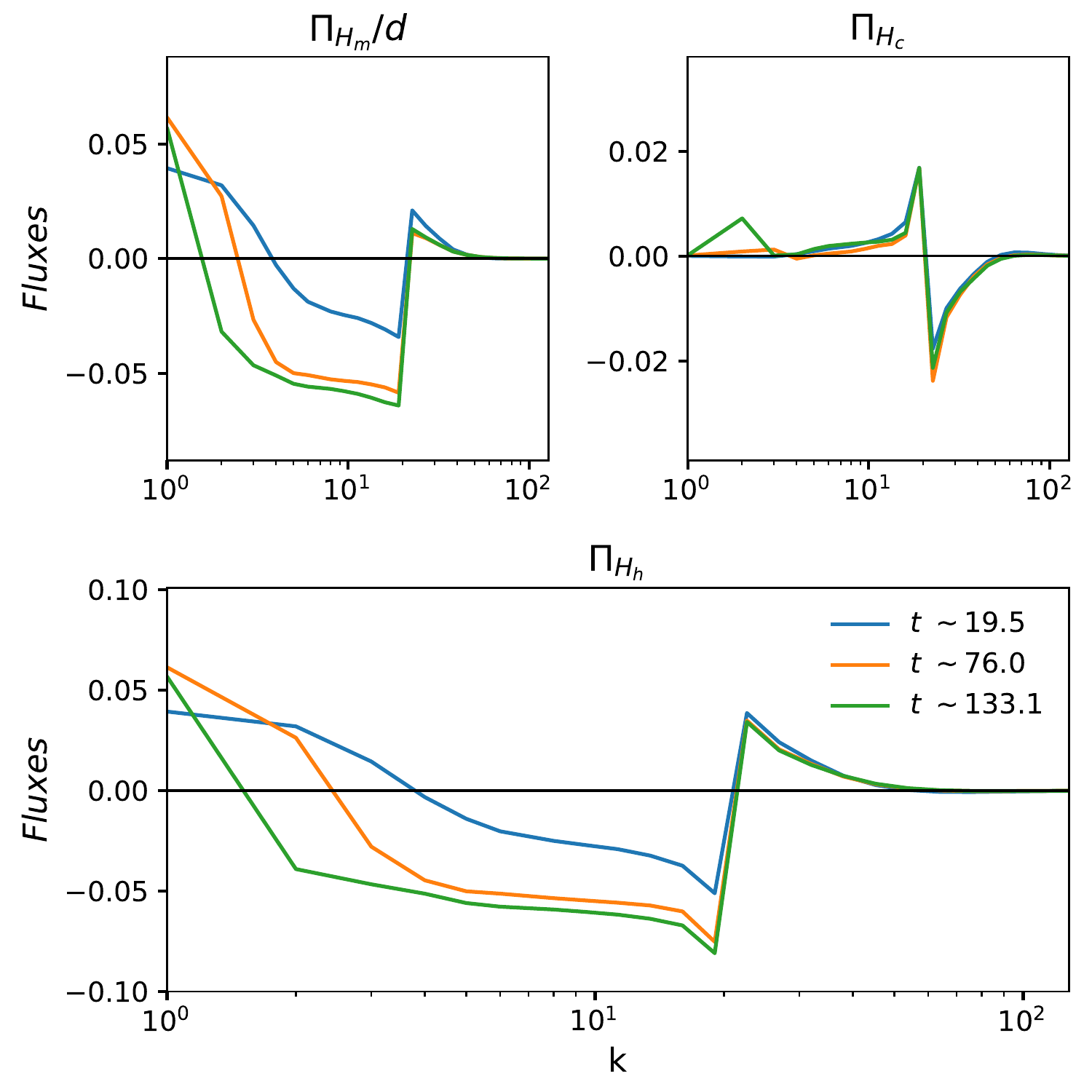}
\caption{Spectral fluxes (in logarithmic-linear scales) of the normalized magnetic helicity (top left) divided by $d$, cross-correlation (top right) and hybrid 
helicity (bottom) for simulation 20R$_0$ at three different times. The hybrid helicity and the magnetic helicity exhibit a plateau at large scale which widens 
with time.}
\label{Fig7}
\end{figure}
To further investigate the dynamics of the hybrid helicity we define the following fluxes 
\ba
\Pi_{H_m} (k) &=&  - \bb_{k}^{<} \cdot (\uu \times \Correction{\bb_{k}^{>} } ) \, , \\
\Pi_{H_c} (k) &=&   \frac{1}{2} [\bb_{k}^{<} \cdot (\uu \cdot \nabla \Correction{ \uu_{k}^{>}} - \bb \cdot \nabla \Correction{\bb_{k}^{>} }) \nonumber \\
&&+ \uu_{k}^{<} \cdot (\uu \cdot \nabla \Correction{\bb_{k}^{>}} - \bb \cdot \nabla \Correction{\uu_{k}^{>} }) ] \, , \\
\Pi_{H_h} (k) &=&   \Pi_{H_m} (k) / d - \Pi_{H_c} (k) \, , 
\ea
for the magnetic helicity, the cross-correlation and the hybrid helicity, respectively. The time evolution of these spectra is shown in Fig. \ref{Fig7} for $\theta=0^o$. 
This figure provides an additional information: the hybrid helicity spectrum (bottom) tends to be formed with a constant negative flux at large scales. 
This negative flux can be attributed to the magnetic helicity (top left) whereas the cross-helicity displays only a slight positive flux (top right). 
It is important to note that unlike total energy, the quantities $H_{m}$, $H_{c}$ and $H_{h}$ are not positive defined. 
However, since the forcing excites preferentially the right fluctuations it is expected to have a positive magnetic helicity (as we checked). 
Since $H_{m}/d$ has a dominant contribution to $H_{h}$ our interpretation about the sign of the hybrid helicity flux is therefore probably correct. 
Note that a positive flux at the largest scales is also observed in other studies and usually interpreted as an effect of the periodic boundaries of the numerical box.

\begin{figure}
\includegraphics[width=1\linewidth]{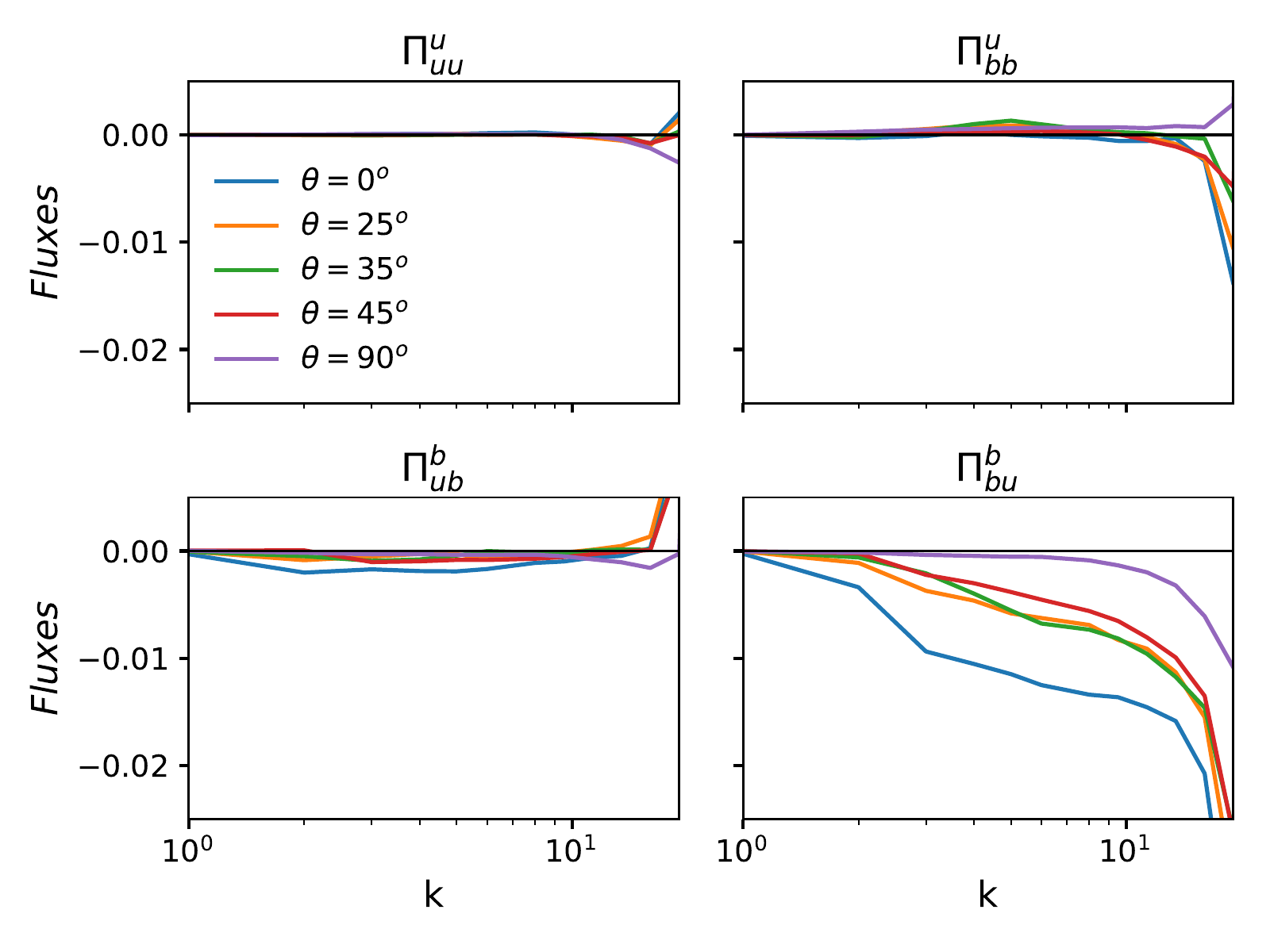}
\caption{Spectral fluxes (in logarithmic-linear scales) associated with the four different nonlinear terms of $B\Omega$-MHD for several angles (simulation $20R_0$ to $20R_{90}$) and at time $t\sim 105$. For the flux $\Pi^b_{bu}$, the difference between $\theta =0^o$ and the other angles is undeniable.}
\label{Fig8}
\end{figure}
In Fig. \ref{Fig8} we see how the angle $\theta$ affects the flux associated with the four different nonlinear terms of $B\Omega$-MHD. 
The most remarkable evolution comes from the bottom-right panel where the main driver of the inverse cascade is plotted: its flux is 
drastically reduced when $\theta$ increases. The inverse transfer is almost completely damped for a large $\theta$ angle. 

\begin{figure}
\includegraphics[width=1\linewidth]{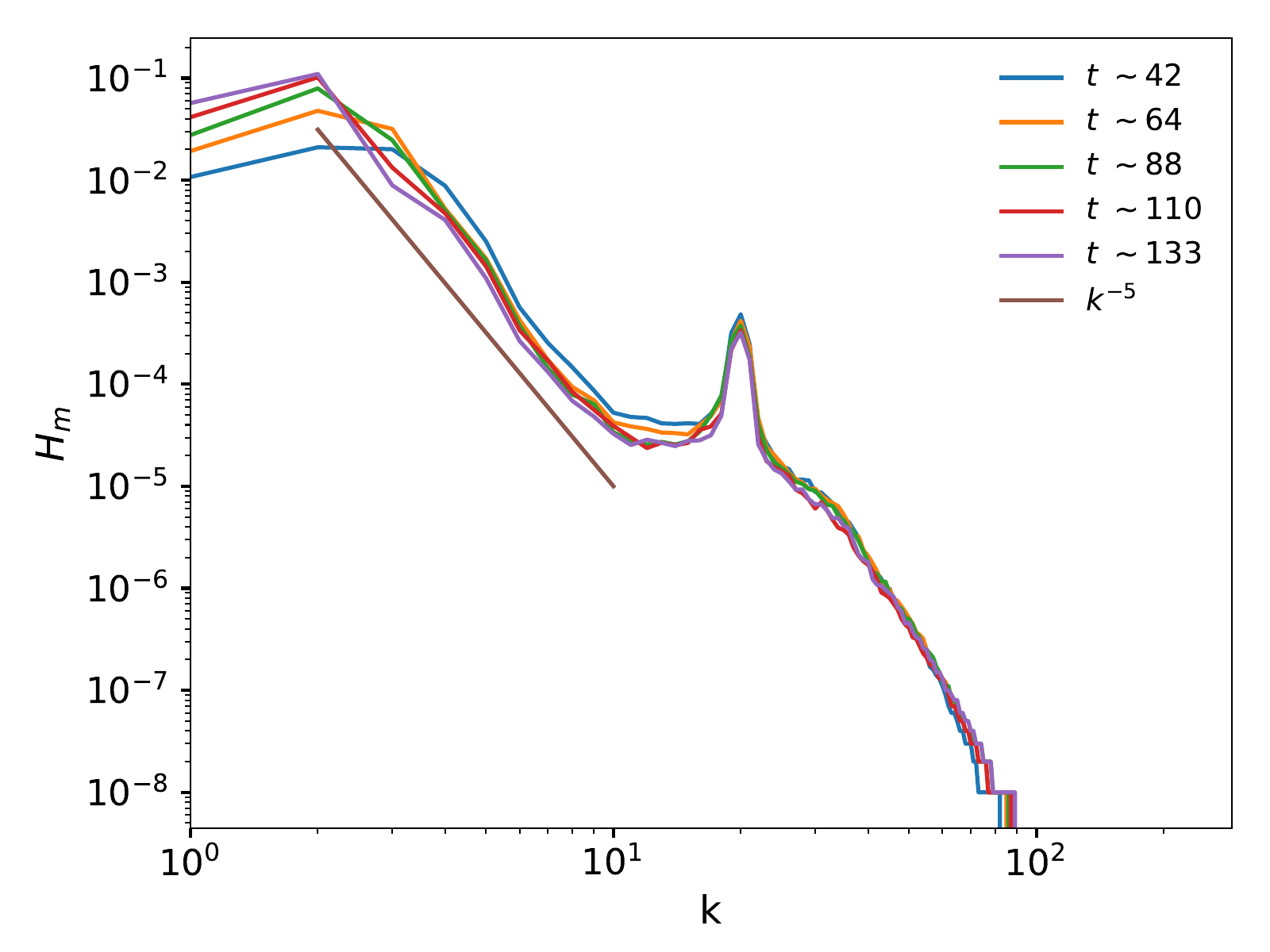}
\caption{Time evolution of the magnetic helicity spectrum for simulation $20R_0$. The spectra are compared with a power law in $k^{-5}$ (segment). 
Note that the spectrum peaks at wavenumber $k=2$ from $t \sim 88$.}
\label{Fig9}
\end{figure}
Finally, the time evolution of the magnetic helicity spectrum is shown in Fig. \ref{Fig9} for $\theta=0^o$ (simulation $20R_0$). As we see the large scale 
spectrum is well fitted with a power law in $k^{-5}$. Even if a power law is found in a narrow wavenumber window we may try to compare it with theoretical 
predictions. The law found is quite different from the pure MHD case ($\Omega_0=B_0=0$) where a direct numerical simulations showed a $k^{-3.6}$ scaling 
\cite{Muller12} or for which a closure model predicted a $k^{-2}$ scaling \cite{pouquet76}. It is also different from the weak wave turbulence prediction 
\cite{Galtier14}. Therefore, the spectrum observed remains unexplained.

%%%%%%%%%%%%%%%%%%%%%
\section{Conclusion}\label{sec4}
%%%%%%%%%%%%%%%%%%%%%

The present study was focused on the impact of a polarized forcing at intermediate scale on the large-scale dynamics of $B\Omega$-MHD turbulence at low
Rossby number. 
The main property found is that a right-handed polarization is much more efficient than a left-handed to excite large-scale magnetic field fluctuations. 
This can be explained by invoking wave turbulence arguments and the hybrid helicity $H_{h}$ which is a conserved quantity when the rotation axis 
$\Ome_{0}$ and the background magnetic field $\bbl$ are aligned. As a consequence of this inviscid property we observe an inverse cascade of $H_{h}$ 
with a constant negative flux. This inverse transfer decreases when the angle $\theta \equiv \widehat{\left(\Ome_{0}, \bbl \right)} > 0^{o}$; it becomes weak 
when $\theta \ge 35^o$. This critical angle is, however, not universal and could be smaller when right- and left-handed fluctuations are equally excited. 

Stars and planets are often characterized by a magnetic dipole closely aligned with the rotation axis. Why\,? The answer to this question is far from 
trivial because it involves many sub-questions linked to the turbulent dynamo problems in a spherical geometry. Usually because of the complexity 
of the problem, only few physical ingredients are included in the modeling like the thermal convection and the rotation which can be simplified by 
considering a solid-body instead of a differential rotation. Furthermore, the magnetic Prandtl number $P_m$ can take very different values if one 
considers stars or planet interiors but generally in both cases $P_m<1$. Last but not least, the conducting fluid is highly turbulent and requires 
power numerical resources if one wants to find solutions that cover a wide range of scales. 

Our study reveals that the regeneration of a large-scale magnetic field can be done through an inverse cascade of hybrid helicity. 
We found that the inverse cascade is more efficient when the angle $\theta$ is small. 
This result is an indication that the dynamo mechanism is more efficient when locally the mean magnetic field is aligned with the rotating rate. 
Generally speaking our study reveals that the hybrid helicity is a fundamental ingredient for the dynamo in $B\Omega$-MHD turbulence at low Rossby number. 

%%%%%%%%%%%%%%%%%%%%%
\appendix
\section{Helicity decomposition}\label{sec5}

The incompressibility conditions ($\ref{rmhd3}$) and ($\ref{rmhd4}$) allow the projection of the $B\Omega$-MHD equations on a complex helicity basis, ie. 
in a plane orthogonal to $\kk$ \cite{Galtier14}. We introduce the complex helicity decomposition
\be
{\bf h^{\Lambda}}(\kk) \equiv \hh = \ete + i \Lambda \efi \, ,
\label{basis0}
\ee
where the wave vector $\kk = k\ek = \kkp + \kpa \ep$ ($k=|\kk|$, $\kpn = |\kkp|$, $|\ek|=1$) and $i^2 = -1$ and where
\ba
\ete &=& \efi \times \ek \, , \\
\efi &=& \frac{\ep \times \ek}{|\ep \times \ek|} \, ,
\label{efi}
\ea
with $|\ete (\kk)|$=$|\efi(\kk)|$=$1$. Note that ($\ek$, $h^+_{\kk}$, $h^-_{\kk}$) form a complex basis with the following properties
\ba
{\bf h^{-\Lambda}_{k}} &=& {\bf h^{\Lambda}_{-k}} \, , \\
\ek \times \hh &=& - i \Lambda \, \hh \, , \\
\kk \cdot \hh &=& 0 \, , \\
{\bf h^{\Lambda}_k} \cdot {\bf h^{\Lp}_k} &=& 2 \, \delta_{-\Lp \Lambda}\, .
\ea
We project the Fourier transform of the original vectors $\uu ({\bf x})$ and $\bb ({\bf x})$ on the helicity basis and find
\ba
\hat \uu_\kk &=& \sum_{\Lambda} \, {\cal U}_{\Lambda} (\kk) \, {\bf h^{\Lambda}_k} 
= \sum_{\Lambda} \, {\cal U}_{\Lambda} \, {\bf h^{\Lambda}_k} \, , \\
\hat \bb_\kk &=& \sum_{\Lambda} \, {\cal B}_{\Lambda} (\kk) \, {\bf h^{\Lambda}_k} 
= \sum_{\Lambda} \, {\cal B}_{\Lambda} \, {\bf h^{\Lambda}_k} \, ,
\label{basis1}
\ea
If we inverse the system, we find the following relations for the velocity components: 
\ba
{\cal U}_+ (\kk) &=& {1 \over 2 k \kpn} [ k_x \kpa \hat u_x + k_y \kpa \hat u_y - \kpn^2 \hat u_z \nonumber \\
&&+ i k (k_y \hat u_x - k_x \hat u_y)] \, ,  \\
{\cal U}_- (\kk) &=& {1 \over 2 k \kpn} [ k_x \kpa \hat u_x + k_y \kpa \hat u_y - \kpn^2 \hat u_z \nonumber \\
&&- i k (k_y \hat u_x - k_x \hat u_y) ] \, . 
\ea
Similar relations are found for the magnetic field. Then, the kinetic and magnetic energy spectra will be given by $\langle \vert {\cal U}_{\Lambda} \vert^2 \rangle$ 
and $\langle \vert {\cal B}_{\Lambda} \vert^2 \rangle$, respectively: for a positive $k_{z}$, $\Lambda=+1$ corresponds to the R-fluctuations and $\Lambda=-1$
to the L-fluctuations.

\section*{Acknowledgments}
This work was supported by the LabEX Pla@par and received financial state aid managed by the Agence National de la Recherche, 
as part of the Programme ``Investissements d'Avenir'' under the reference ANR-11-IDEX-0004-02. 
This work was granted access to the HPC resources of CINES under allocation 2017 A0030410072 made by GENCI.
We thank R. Meyrand for useful discussions. 

\bibliography{geophys} 

\end{document}